# NMR diffusion pore imaging: Experimental phase detection by double diffusion encoding

Kerstin Demberg[1], Frederik Bernd Laun[1,2], Johannes Windschuh[1], Reiner Umathum[1], Peter Bachert[1] and Tristan Anselm Kuder[1]*

[1] *Medical Physics in Radiology, German Cancer Research Center (DKFZ), Heidelberg, Germany*

[2] *Institute of Radiology, University Hospital Erlangen, Erlangen, Germany*

**\*Address of correspondence:**

Department Medical Physics in Radiology

German Cancer Research Center (DKFZ)

Im Neuenheimer Feld 280

D-69120 Heidelberg

Phone: +49 6221 42 2410

Fax: +49 6221 42 2585

Email: t.kuder@dkfz.de

**Abstract**

Diffusion pore imaging is an extension of diffusion-weighted nuclear magnetic resonance imaging enabling the direct measurement of the shape of arbitrarily formed, closed pores by probing diffusion restrictions using the motion of spin-bearing particles. Examples of such pores comprise cells in biological tissue or oil containing cavities in porous rocks. All pores contained in the measurement volume contribute to one reconstructed image, which reduces the problem of vanishing signal at increasing resolution present in conventional magnetic resonance imaging. It has been previously experimentally demonstrated that pore imaging using a combination of a long and a narrow magnetic field gradient pulse is feasible. In this work, an experimental verification is presented showing that pores can be imaged using short gradient pulses only. Experiments were carried out using hyperpolarized xenon gas in well-defined pores. The phase required for pore image reconstruction was retrieved from double diffusion encoded (DDE) measurements, while the magnitude could either be obtained from DDE signals or classical diffusion measurements with single encoding. The occurring image artifacts caused by restrictions of the gradient system, insufficient diffusion time, and by the phase reconstruction approach were investigated. Employing short gradient pulses only is advantageous compared to the initial long-narrow approach due to a more flexible sequence design when omitting the long gradient and due to faster convergence to the diffusion long-time limit, which may enable application to larger pores.

# I. Introduction

Diffusion-weighted nuclear magnetic resonance (NMR) imaging allows probing diffusion restrictions such as cell membranes in biological tissues [1, 2] or porous structures in rocks [3, 4] by measuring the diffusive motion of water molecules. This way, parameters that are indirectly linked to the microstructure of tissues or porous media are derived, such as the apparent diffusion coefficient or the fractional anisotropy characterizing the degree of directionality of the diffusive motion. Most prominent medical applications are the diagnosis of ischemic stroke and tumor detection [1, 5-10]. Furthermore, nerve fiber connections in the human brain can be visualized by tracking white matter pathways [11, 12]. However, in contrast to those parameters indirectly related to the structure of diffusion restrictions, it would be highly desirable to obtain measures directly characterizing the confining geometries, which will be denoted as pores in this work. This would enable the direct measurement of pore sizes and shapes or the in vivo estimation of histology-like parameters such as cell size distributions. A classical approach in this direction is the so-called q-space imaging technique [13, 14], which uses two short magnetic field gradient pulses for diffusion encoding. However, this approach only allows measuring the power spectrum of the pore space function. Due to the loss of the phase information, an unambiguous reconstruction of closed pores filled with an NMR-detectable diffusing medium is not possible.

More recently, Laun et al. proposed a modified temporal gradient profile for diffusion encoding which uses a combination of a long and a short gradient pulse [15]. Using this gradient profile, the phase information is preserved enabling the direct reconstruction of arbitrary pore space functions. If many identical pores are contained in the considered volume element, all pores contribute to a single pore image. Due to this averaging effect, the problem of vanishing signal when enhancing the image resolution is vastly reduced compared to conventional magnetic resonance imaging (MRI). In principle, this diffusion pore imaging approach could be employed to measure cell or pore size and shape distributions [16, 17]. Using this long-narrow gradient approach, it was experimentally proven that average images of closed pores can be obtained [18-20]. However, the use of very long gradient pulses restricts the applicable measurement sequence techniques considerably. Therefore, using short gradient pulses only would be desirable, since this would allow the application of a broader spectrum of measurement sequences, for example, using stimulated echoes.

It has been previously noted, that signals acquired using three short gradient pulses, a special case of double diffusion encoded (DDE) measurements [21], can exhibit imaginary signal parts [22] and thus contain phase information. A method using q-space and DDE experiments to reconstruct point-symmetric pore shapes has been proposed and demonstrated experimentally using cylindrical capillaries [23]. Subsequently, methods using only short gradient pulses for reconstruction of arbitrary pore shapes were proposed [24]. These methods were then generalized to a wider class of temporal gradient profiles [25].

The objective of this work was to provide a first experimental demonstration of the possibility to image non-point-symmetric pore shapes using short-gradient methods. For this purpose, experiments in well-defined pores were conducted on a clinical MR-scanner and two reconstruction methods described in [24] were used. The obtained pore images were compared to measurements using the long-narrow approach.

## II. Theory

When studying porous structures, a closed pore is described by the pore space function $\rho(x)$. Inside the pore, $\rho(x)$ equals the reciprocal of the volume $V$ of the pore, while $\rho(x)$ is zero outside the pore. Using pore imaging techniques, the Fourier transform $\tilde{\rho}(q)$ of $\rho(x)$ is measured, where $q$ denotes a vector in q-space.

In the following, the three methods used here for the experimental demonstration are summarized. Method 1 and 2 use short gradient pulses only, while method 3 denotes the long-narrow approach used for comparison.

*Method 1: Phase estimation by combination of q-space and DDE measurements*

For single diffusion encoded q-space measurements, two short diffusion gradients with opposite polarity are applied. This gradient profile is depicted in Fig. 1(a), where a 180° radiofrequency (RF) pulse was inserted to realize a spin echo sequence. The first gradient pulse dephases the magnetization, while the second one results in a rephasing. In the presence of diffusive motion the rephasing is incomplete, causing a signal attenuation. The acquired spin phase in the rotating reference frame is given by $\varphi = -\gamma \int_0^T \boldsymbol{G}(t) \cdot \boldsymbol{x}(t) dt$ where $x(t)$ denotes the spin position at time $0 < t < T$, $\boldsymbol{G}(t)$ the temporal gradient profile and $\gamma$ the gyromagnetic ratio [26].

The two gradient pulses with the gradient vectors $-\boldsymbol{G}$ and $\boldsymbol{G}$ and duration $\delta$ are applied at $t = 0$ and $t = T - \delta$. This definition satisfies the rephasing condition resulting in a vanishing zeroth gradient moment, and ensures that immobile spins do not accumulate a net phase. Introducing the q-vector $\boldsymbol{q} = \gamma \boldsymbol{G} \delta$, the echo amplitude attenuation for the average over all random walks is given by

$$S_{11}(\boldsymbol{q}) = \langle \exp(i\varphi) \rangle = \left\langle \exp\left[-i\boldsymbol{q}\left(-\frac{1}{\delta}\int_0^\delta x(t)dt + \frac{1}{\delta}\int_{T-\delta}^T x(t)dt\right)\right]\right\rangle. \tag{1}$$

The magnitude of the vector $\boldsymbol{q}$ is referred to as the q-value. The integrals $\frac{1}{\delta}\int x(t)\,dt$ in Eq. (1) equal the centers of mass of the random walk during the application of the first gradient pulse ($\boldsymbol{x}_{\text{cm},1}$) and the second one ($\boldsymbol{x}_{\text{cm},2}$). Thus, the phase acquired by a particle is the same as for a particle resting at the center of mass of the respective trajectory [26]. Consequently, Eq. (1) can be rewritten to $S_{11}(\boldsymbol{q}) = \langle \exp[i\boldsymbol{q}(\boldsymbol{x}_{\text{cm},1} - \boldsymbol{x}_{\text{cm},2})]\rangle$. In the limit of short gradient pulses ($\delta \to 0$) and long diffusion time ($T \to \infty$), the signal becomes [13]

$$S_{11}(\boldsymbol{q}) = \langle e^{-iq(x_2-x_1)} \rangle = \langle e^{iqx_1} \rangle \langle e^{-iqx_2} \rangle = \int_{\text{Pore}} \rho(x_1) e^{iqx_1} dx_1 \int_{\text{Pore}} \rho(x_2) e^{-iqx_2} dx_2$$

$$= \tilde{\rho}^*(\boldsymbol{q})\tilde{\rho}(\boldsymbol{q}) = |\tilde{\rho}(\boldsymbol{q})|^2, \tag{2}$$

where $x_1 = x(0)$ and $x_2 = x(T)$ denote the initial and final position of the respective random walk. The complex factor $e^{-iq(x_2-x_1)}$ contains the phase that is accumulated during the particle displacement $x_2 - x_1$. The precondition $T \to \infty$ of this equation ensures that initial and final position of each random walk are uncorrelated so that averaging independently over $x_1$ and $x_2$ becomes possible. From Eq. (2), only the magnitude of $\tilde{\rho}(\boldsymbol{q})$ can be obtained, but the phase information is lost. This means that $\rho(x)$ cannot be determined unambiguously from q-space measurements alone. This predicament is found analogously in X-ray and neutron scattering experiments, where it is commonly referred to as the "phase problem" [27] and where $\tilde{\rho}(\boldsymbol{q})$ is called the "form factor".

While all phase information is lost in q-space imaging, applying its gradient profile twice in succession, thus using two diffusion encodings, results in signals containing phase information [22]. For this type of experiment, the term double diffusion encoding (DDE) is widely accepted [21].

For the DDE gradient profile considered here, the two diffusion encodings are applied with anti-parallel wave vectors without temporal separation. The second and third gradient pulse are superimposed, which results in three short gradient pulses of duration $\delta$ with the following gradient vectors: $-\boldsymbol{G}/2$ starting at $t = 0$ and $t = T - \delta$, and $\boldsymbol{G}$ starting at $t = T/2 - \delta/2$. This gradient profile is depicted in Fig. 1(b), also extended by a 180° RF pulse. Under idealized conditions, where short gradient pulses ($\delta \to 0$) and long diffusion time in each interval ($T \to \infty$) are assumed, the signal drop is given by [22, 23]

$$S_{121}(\boldsymbol{q}) = \tilde{\rho}^*(\boldsymbol{q}/2)^2 \tilde{\rho}(\boldsymbol{q}). \tag{3}$$

For non-point-symmetric pores, $\tilde{\rho}(\boldsymbol{q})$ is complex, for point-symmetric pores $\tilde{\rho}(\boldsymbol{q})$ is real, where positive and negative values can occur [16, 22, 24]. $S_{121}(\boldsymbol{q})$ clearly contains phase information of $\tilde{\rho}(\boldsymbol{q})$, but it is mixed for different q-values. To extract the phase information from $S_{121}(\boldsymbol{q})$, an iterative approach was proposed in Ref. [24], where $\tilde{\rho}(\boldsymbol{q})$ and $S_{121}(\boldsymbol{q})$ are written in terms of magnitude and phase:

$$\tilde{\rho}(\boldsymbol{q}) = A(\boldsymbol{q}) e^{i\psi(\boldsymbol{q})} \text{ and } S_{121}(\boldsymbol{q}) = B(\boldsymbol{q}) e^{i\theta(\boldsymbol{q})}, \tag{4}$$

where $A(\boldsymbol{q})$ can be derived from $S_{11}(\boldsymbol{q})$ using Eq. (2), i.e. $A(\boldsymbol{q}) = \sqrt{S_{11}(\boldsymbol{q})}$. To find $\psi(\boldsymbol{q})$, the expressions for $S_{121}(\boldsymbol{q})$ and $\tilde{\rho}(\boldsymbol{q})$ are inserted in Eq. (3):

$$B(\boldsymbol{q}) e^{i\theta(\boldsymbol{q})} = A(\boldsymbol{q}/2)^2 \, e^{-2i\psi(\boldsymbol{q}/2)} A(\boldsymbol{q}) e^{i\psi(\boldsymbol{q})}. \tag{5}$$

The magnitudes have to be real and positive, which allows comparing the exponential factors, obtaining a recursive formula to iteratively estimate the phase $\psi(\boldsymbol{q})$ of $\tilde{\rho}(\boldsymbol{q})$ from the phase $\theta(\boldsymbol{q})$ of $S_{121}(\boldsymbol{q})$:

$$\psi(\boldsymbol{q}) = 2\psi(\boldsymbol{q}/2) + \theta(\boldsymbol{q}). \tag{6}$$

*Method 2: Magnitude estimation using DDE measurements*

It was shown that the information contained in the DDE measurement is theoretically sufficient to reconstruct the pore space function $\rho(\boldsymbol{x})$ without obtaining the magnitude from $S_{11}(q)$ [24]. Therefore, the q-space measurement can be skipped. This approach is also verified experimentally in this work. The magnitudes of the complex numbers in Eq. (5) have to be equal on both sides leading to

$$A(\boldsymbol{q}) = \frac{B(\boldsymbol{q})}{A(\boldsymbol{q}/2)^2}. \tag{7}$$

*Method 3: Long-narrow approach*

For comparison, measurements using the long-narrow approach were performed [15]. In recent experiments, this method has already been demonstrated to be capable of measuring average images of cylindrical and triangular pores [18-20]. For this method, a long-narrow gradient profile is used, which is non-antisymmetric in time. The first gradient pulse in q-space imaging is replaced by a gradient of increased duration $\delta_\text{L}$ while the duration of the second gradient remains short ($\delta_\text{N}$). The long gradient is applied with reduced amplitude $G_\text{L}$ to satisfy the condition $\boldsymbol{G}_\text{L}\delta_\text{L} = -\boldsymbol{G}_\text{N}\delta_\text{N}$. In contrast to q-space imaging, the phase information is preserved. The position of the short gradient within the temporal gradient profile is irrelevant in the limit of long diffusion time and short gradient duration. A possible long-narrow gradient profile is depicted in Fig. 1(c). Here, the short gradient pulse was split into two parts to insert a 180° RF pulse. This is possible if the typical diffusion distance of the particles during the separating interval is short enough.

The signal obtained when using a long-narrow gradient profile can again be expressed using the centers of mass $\boldsymbol{x}_\text{cm,L}$ and $\boldsymbol{x}_\text{cm,N}$ of the particle trajectories during the long and short gradient pulse analogously to the derivation in the q-space imaging section [15, 26]:

$$S_{\text{LN}}(\boldsymbol{q}) = \langle \exp[i\boldsymbol{q}(\boldsymbol{x}_{\text{cm,L}} - \boldsymbol{x}_{\text{cm,N}})]\rangle. \tag{8}$$

In the long-time limit ($T \to \infty$), the particles have explored the whole pore so that $\boldsymbol{x}_{\text{cm,L}}$ converges to the pore center of mass $\boldsymbol{x}_{\text{cm}}$. For $\delta_{\text{N}} \to 0$, $\boldsymbol{x}_{\text{cm,N}}$ converges to a single trajectory point. Therefore, the term for the narrow pulse results in the form factor:

$$S_{\text{LN}}(\boldsymbol{q}) = \exp[i\boldsymbol{q}\boldsymbol{x}_{\text{cm}}]\tilde{\rho}(\boldsymbol{q}). \tag{9}$$

The phase factor $\exp[i\boldsymbol{q}\boldsymbol{x}_{\text{cm}}]$ results from the long gradient pulse and ensures that – in the case of many pores in the volume element – all pores are superimposed and an average pore image is formed [15, 17]. Pore images are obtained directly by taking the inverse Fourier transform of the signal.

## III. Experiment

We performed all experiments on phantoms utilizing xenon-129 gas as the diffusing medium inside the pores. Due to the weak signal of thermally polarized xenon gas, hyperpolarized gas with a considerably higher NMR signal was employed. A gas mixture (Air Liquide Deutschland GmbH, Düsseldorf, Germany) composed of xenon (0.95 Vol.-%), nitrogen (8.75 Vol.-%) and helium (rest) was utilized. Owing to the large gas diffusion coefficient, it was possible to reach the diffusion long-time limit using phantoms with pores on the millimeter scale. The free diffusion coefficient of the hyperpolarized xenon gas mixture was estimated to $D_0$ = (37000±2000) µm$^2$/ms [18], which is one order of magnitude larger than for pure xenon gas [28].

Hyperpolarization was achieved by Rb/Xe-129 spin-exchange optical pumping (SEOP) [29], a process involving two main steps: First, optical pumping of rubidium atoms to generate an electron-spin polarization and second, the spin transfer from the rubidium electron to the Xe-129 nucleus. External heating was used to generate rubidium vapor in a pumping cell which was located in a pair of Helmholtz coils to generate a static magnetic field. The vapor mixed with the gas mixture that was flowing into the pumping cell, which was operated at a pressure of 1 bar above atmospheric pressure. A high helium concentration was used to pressure broaden the Rb $D_1$ line to reach better laser light absorption and nitrogen was added to quench the excited state of the Rb atoms. Light from a diode laser (Coherent FAP Duo, Coherent Inc., Santa Clara, CA, USA, 60 W, 795 nm) was right-circularly polarized and was irradiating the pumping cell. After the hyperpolarized gas left the pumping cell, its pressure was reduced to atmospheric pressure and it was transferred to a diffusion pore imaging phantom via a polyurethane pipe.

Two types of phantoms were studied: The phantom type I consisted of acrylic glass plates with groves of equilateral triangular shape, which were stacked on top of each other built by the in-house mechanical workshop [Fig. 1(d)]. One set consisted of 170 triangles with an edge length of $L$ = 3400 µm and a second set contained 77 triangles with $L$ = 5750 µm. For the phantom type II [Fig. 1(e)], three blocks containing tubes with different cross-sectional shapes were 3D-printed using the PolyJet technology (Objet30 Pro, VeroClear as printing material, Stratasys, Ltd., Eden Prairie, MN, USA). The following cross-sectional shapes were used: "Half-moon" (210 pores, $L$ = 3450 µm), "fish" (165 pores, $L_1$ = 3450 µm, $L_2$ = 2350 µm) and "star" (135 pores, $L_1$ = 3100 µm, $L_2$ = 3000 µm). All experiments were performed on a clinical MR scanner at $B_0$ = 1.5 T (Magnetom Symphony, A Tim System, Siemens Healthcare, Erlangen, Germany) and the maximum gradient amplitude used was 32 mT/m. Each phantom was placed at the isocenter of the scanner with the gas flow of approximately 140 ml/min through the pores parallel to the main magnetic field.

The radiofrequency pulses were applied slice-selectively (slice thickness 35 mm). The normal vector of the slice plane was parallel to the flow direction, so that the diffusion encoding gradients were applied in the transversal plane of the scanner orthogonal to the gas flow direction. The q-space was sampled two-dimensionally and radially using the gradient profiles depicted in Fig. 1 (a)-(c), where 180° refocusing pulses with a duration of 2.048 ms were inserted in all gradient profiles to realize spin echo sequences. The 180° pulses were surrounded by spoiler gradients in slice selection direction. For each gradient direction, the gradient amplitude was gradually increased while keeping the gradient timing constant and the spin echo signal was recorded. The time difference between two 90° excitation pulses was set to 18 seconds to achieve a sufficient gas exchange in the phantom to restore the polarization before a new point in q-space was acquired. An additional signal acquisition directly after the 90° excitation pulse was used to normalize the recorded spin echo amplitudes to account for fluctuations in the polarization level. All gradient durations $\delta$, $\delta_N$ and $\delta_L$ are given as ramp up time plus flat top time of the trapezoidal gradient pulses. Ramp up times were 0.32 ms for all short gradient pulses and 0.16 ms for the long gradient pulses in Fig. 1(c). The signal values of each spin echo readout were averaged and normalized to the pre-readout signal. To obtain the diffusion-induced signal attenuation, additional normalization to $S(\boldsymbol{q}=0)$ was conducted.

One-dimensional pore imaging experiments with the triangle phantom ($L$ = 3400 μm) for two orthogonal gradient directions (bottom-top and left-right in Fig. 1(d)) were performed using the q-space and the DDE gradient profile with 18 q-values. To generate two-dimensional pore images, q-space was sampled radially with 37 gradient directions distributed in 5° steps in one half-space, while the second half-space was calculated by $S(-\boldsymbol{q}) = S^*(\boldsymbol{q})$. This assumption was verified experimentally for three gradient directions for all pore shapes in preparatory experiments.

For the short-gradient measurements, method 1 or 2 was used to estimate $\tilde{\rho}(\boldsymbol{q})$ for each gradient direction. For method 1, Eq. (6) was solved recursively as described in [24]: As initial conditions, $\psi(\boldsymbol{q})$ and its derivative were set to zero for $\boldsymbol{q} = 0$. Thus, radial acquisitions of $S_{121}(\boldsymbol{q})$ in q-space can be used to reconstruct the phase $\psi(\boldsymbol{q})$. $\psi(\boldsymbol{q}_1) = 0$ was assumed for the first non-zero q-value $q_1$. $\psi(\boldsymbol{q})$ was then recursively reconstructed for higher q-values using appropriate interpolation, i.e. if $\boldsymbol{q}/2$ in Eq. (6) did not coincide with a measured value, the respective value $\psi(\boldsymbol{q}/2)$ needed for the recursive reconstruction was obtained by linear interpolation. For method 2, Eq. (7) is used together with $1 - A(\boldsymbol{q}) \propto \boldsymbol{q}^2$ for small $q$ to estimate $A(\boldsymbol{q})$ recursively [24]. Values for $A(\boldsymbol{q}/2)$ in Eq. (7) were obtained analogously to the interpolation process of method 1. As described in [24], a threshold for the magnitude was introduced in method 2 to prevent erroneous reconstruction near zero crossings: For points in gradient direction $\boldsymbol{n}$ with q-values $q = |\boldsymbol{q}|$, we assume, that if $A(\boldsymbol{n}q_{\text{th}}) < 0.08$ for a specific value $q_{\text{th}}$, then $A(\boldsymbol{n}q) < 0.22$ for all $q > q_{\text{th}}$. If this condition was not met, $A(\boldsymbol{n}q)$ was set to zero.

Afterwards, an inverse Fourier transform was applied for each direction returning the projection of the pores onto the gradient direction. For the long-narrow gradient profile, the Fourier transform was applied directly to each measured radial line in q-space. To reconstruct two-dimensional images, an inverse Radon transform was applied. Signal processing and image reconstruction were performed in Matlab (MathWorks, Natick, MA, USA).

For simulations of the diffusion process in the triangular pores, an eigenvalue decomposition approach as was used (Eq. (144) of [30]), see also [16, 31-34]: To solve the Bloch-Torrey equation, the transversal magnetization is expanded in the basis of precalculated eigenfunctions of the Laplace operator for the respective pore shape. This results in a set of ordinary differential equations for the diffusion encoded signal, which can be written in matrix representation. Due to the fast decay of contributions of terms with large eigenvalues, using a limited number of eigenfunctions is sufficient to get a good approximation of the signal at relatively low computational time. For the triangular domain, the necessary Laplacian eigenfunctions and the matrices are known analytically [16]. Monte Carlo simulations with $1.5 \cdot 10^6$ random walkers and $6 \cdot 10^4$ steps per random walk trajectory were used for the more complex pore shapes (Fig. 1(e)) to study the signal behavior prior to the image acquisition. In both cases, the computation of the diffusion-encoded signal was implemented in Matlab.

## IV. Results

First, we consider method 1 (phase estimation) that combines the q-space and the DDE measurement. Each row of Fig. 2 shows radial q-space ($S_{11}(\mathbf{q})$) and DDE ($S_{121}(\mathbf{q})$) acquisitions of the equilateral triangles ($L$ = 3400 µm) using the indicated gradient directions, as well as the estimated form factor $\tilde{\rho}(q)$ in q-space, and its Fourier transform $\rho(x)$ into x-space. The simulations are indicated by lines and the experimental results by symbols. As expected, the q-space imaging signal only provides magnitude information, i.e. the measured values of the imaginary part of $S_{11}(\mathbf{q})$ all lie very close to zero. But, for this non-point-symmetric pore shape, an imaginary signal part arises in the DDE signal, which contains the full phase information and allows determining $\tilde{\rho}(q)$ [Fig. 2(a)]. The one-dimensional inverse Fourier transform of the radial acquisition of $\tilde{\rho}(q)$ with vertical gradient direction clearly shows the projection of the pores onto this direction. This one-dimensional x-space profile is composed of the contributions of all pores in the imaging volume. For the horizontal gradient direction, the projection onto the gradient direction is mirror-symmetric resulting in a purely real DDE signal [Fig. 2(b)] (see section V. of [16]). For both vertical and horizontal gradient direction, the experimental results are in good agreement with the respective simulations for $S_{11}(q)$, $S_{121}(q)$, $\tilde{\rho}(q)$ and $\rho(x)$. Small deviations can, however, be observed at high q-values for $\tilde{\rho}(q)$, which is mainly caused by the small signal values of $S_{11}(q)$ and $S_{121}(q)$.

In Figs. 2(c) to (e), the effect of varying the gradient timing is depicted. When increasing the gradient pulse duration to $\delta = 10.7$ ms or $\delta = 35.8$ ms [Figs. 2(c), (d)], the pore size in the reconstructed projection $\rho(x)$ shrinks. To derive $S_{11}(\mathbf{q})$ and $S_{121}(\mathbf{q})$ in Eq. (2) and (3), it was assumed that the diffusive motion during the application of the gradient pulses is negligible, so that the center of mass of the particle trajectory part during the gradient application equals a single trajectory point. When prolonging $\delta$, this is no longer the case. This makes no difference for positions in the bulk of the pore domain. However, for particles diffusing near the pore boundary during gradient application, the respective trajectory center of mass will be shifted towards the center of the pore resulting in a reduced reconstructed pore size with a signal increase near the boundary. This edge enhancement effect [16, 35] grows with increasing gradient duration [Fig. 2(c) and 2(d)]. In Fig. 2(e), $T$ is reduced from 270 ms to 50 ms and is too short for the random walkers to travel through the whole pore. The long diffusion time requirement is a precondition of Eq. (2) an (3) and when not met, the pore shape gets lost.

Figure 3 shows one-dimensional reconstruction results for method 2, where both magnitude and phase are recursively estimated from the DDE experiment only. The DDE signal $S_{121}(q)$ depicted in Fig. 3 is the same signal as displayed in Fig. 2(a). In Fig. 3, however, the magnitude $A(q)$ of $\tilde{\rho}(q)$ was estimated from the magnitude $B(q)$ of the DDE signal to find the form factor $\tilde{\rho}(q)$ in q-space [Eq. (7)]. At large q-values (approx. $q$ > 6 mm$^{-1}$), the form factor magnitude $A(q)$ for the experiment and the simulation start to differ considerably. At these q-values, the DDE signal has dropped almost to zero which is then divided by $A(q/2)^2$, so that errors in the small values of $A(q/2)^2$ get amplified in the estimation of $A(q)$. Additionally, inaccuracies in the phase estimation occur. The projection of the pore shape, i.e. $\rho(x)$, is nonetheless depicted in a decent quality.

In Fig. 4, measured pore images of the equilateral triangles ($L$ = 3400 µm) are compared to simulations. A pore image using method 1, which utilizes short gradient pulses only ($G_{11}(t)$ and $G_{121}(t)$), is compared to a pore image acquired with method 3 using the long-narrow gradient profile $G_{\text{LN}}(t)$ [Fig. 4(a),(f)]. Both were imaged with the same resolution, i.e. the largest q-value was set to the same value for both images. In both cases, the pore shape is clearly observable. But the pore image generated using method 3 ($G_{\text{LN}}(t)$) shows considerable blurring and the pore size appears reduced, which can be attributed to the finite duration $\delta_{\text{L}} = 279$ ms of the long gradient pulse. This blurring is worse than for method 1 and 2 due to the general slower convergence of method 3 towards the long-time limit. Figure 4 also shows pore images acquired using method 1 with prolonged $\delta$ resulting in strong edge enhancement, and thus a shrinkage occurs. The signal intensity is shifted from the boundary to the center of the domain [Fig. 4(b),(c)]. Shortening of $T$ also leads to a size reduction of the reconstructed pore image [Fig. 4(d)]. Figure 4(e) shows a pore image obtained by method 2, where the magnitude of the form factor was estimated from $S_{121}(\boldsymbol{q})$ using the same signal measurement as in Fig. 4(a). The results of both methods agree very well. The increased sensitivity of method 2 to noise is visible, but does not prevent the revealing of the pore shape.

Figure 5 shows pore image measurements as well as simulations for triangular pores with a larger edge length ($L = 5750$ µm). The pore image obtained by means of the short-gradient approach using phase estimation (method 1) is much more affected by noise compared to the long narrow approach (method 3). Due to the longer diffusion time compared to the measurements in Fig. 4, the overall signal is lower as a result of T2-relaxation. Since the signals $S_{11}(q)$ and $S_{121}(q)$ decay faster with increasing $q$ than $S_{\text{LN}}(q)$, method 1 is more sensitive to noise than method 3. Method 2 (magnitude estimation) is inherently even stronger affected by noise, which prevents the reconstruction of the pore shape in case of Fig. 5(b). At the same time, unlike for the other methods, the diffusion time for the long-narrow profile is still too short resulting in a prominent blurring [Fig. 5(c)].

Figure 6 demonstrates that pore imaging using the short-gradient methods 1 and 2 is possible for arbitrary pore shapes using three examples. The half-moon pore shape is clearly visible for both methods. For the fish-shaped pore, imaging the tail-fin is most problematic due to the edge enhancement effect, which pushes the signal out of the narrow structure. For method 2, the increased sensitivity to noise results in a distortion of the fish shape. For the star-shaped pore, a higher maximal q-value is necessary to resolve the very small angles of the star. Due to limited SNR at the higher q-values, the q-space and DDE signals were acquired four times for the star-shaped pores and then averaged before the reconstruction. For method 1, the pore shape is clearly visible, for method 2 the angles are blurred and less distinguishable from the bulk of the star. In comparison, the pore shapes acquired with method 3 appear blurred and shrunken in size (right column in Fig.(6)). In contrast to Fig. 4 and 5 the simulations exhibit noise because the Monte Carlo method was used instead of the eigenvalue decomposition approach.

## V. Discussion

In this work, a first experimental demonstration of imaging non-point-symmetric pores using the complex signal obtained by double diffusion encoded measurements could be presented. Diffusion pore imaging was feasible for arbitrary pore shapes by double diffusion encoding in combination with q-space measurements, as well as without the information from the latter. All pores in the imaging volume contribute to one reconstructed pore image, so that the signal-to-noise ratio (SNR) limitation of the resolution of conventional NMR imaging no longer applies.

Comparing the results of the short-gradient methods 1 and 2 with the long-narrow approach (method 3) confirms the increased sensitivity of the short-gradient methods to signal noise. Regarding method 1, $S_{11}(\boldsymbol{q})$ decays much faster with increasing q-value, compared to the long-narrow approach [16]. That way, the oscillations of $\tilde{\rho}(\boldsymbol{q})$ at large q-values get lost since taking in the square-root $|\tilde{\rho}(\boldsymbol{q})| = \sqrt{S_{11}(\boldsymbol{q})}$ amplifies the noise in $S_{11}(\boldsymbol{q})$ for small signal values. The DDE signal decay, whose phase is denoted by $\theta(\boldsymbol{q})$, is also fast and thus the input $\theta(\boldsymbol{q})$ to the phase estimation is corrupted at large q-values due to the low SNR. In addition, the phase estimation is impaired at large q-values by the recursive use of phase values at smaller ones, i.e. the reconstructed phases $\psi(\boldsymbol{q}/2)$ and indirectly $\theta(\boldsymbol{q})$ at small q-values. This way, errors in $\theta(\boldsymbol{q})$ at small q-values may be amplified to larger errors in $\psi(\boldsymbol{q})$ at higher q-values. For method 2, the error amplification in the estimation of the signal magnitude additionally yields an increased influence of noise, which is especially problematic near zero crossings of $A(\boldsymbol{q}/2)$ in Eq. (7). Global reconstruction approaches may reduce these stability problems and could take the information from nearby directions into account.

Further, the radial acquisition and reconstruction is not optimal regarding the q-space sampling and in terms of utilizing all information that is available. In the image domain, the pore image is very sparse in the case of a single pore shape and size, because only the pore boundary needs to be determined in this case. Much more points in q-space are sampled compared to the number of pointes needed to characterize the boundary of a single pore shape. Therefore, compressed sensing-like reconstruction approaches may largely improve the image quality or reduce measurement time. Since most of the information is contained in the higher q-values, the acquisition could be improved by scaling the sampling density with a power of distance from the q-space origin.

Regarding the gradient timing, mainly the same requirements could be observed experimentally for the short-gradient methods as for the long-narrow approach, namely long diffusion time and high gradient amplitude to realize sufficiently short gradient pulses. If the short gradient pulses have to be prolonged due to insufficient gradient amplitude, the pore shape will be obscured due to the edge enhancement effect, which occurs for all pore imaging methods. This effect limits the reachable image resolution, since an increase of $q_{\mathrm{max}}$ also results in prolonged $\delta$ for a given maximal gradient amplitude causing small structures to be unobservable, which stresses the high requirements of pore imaging on the gradient system.

Regarding the effect of insufficient diffusion time $T$, a more pronounced blurring occurs for the long-narrow approach (method 3, Fig. 4(f)). For method 3, these artifacts can be attributed to the finite duration of the long gradient pulses. $x_{\mathrm{cm,L}}$ is not equal to the center of mass of the pore $x_{\mathrm{cm}}$ as it would be the case for infinite $T$, but it follows a Gaussian distribution around $x_{\mathrm{cm}}$ [16]. To reach the needed convergence of $x_{\mathrm{cm,L}}$ to $x_{\mathrm{cm}}$, all particles have to explore the whole pore. For method 1, the demand on the duration of $T$ is reduced. $T$ still has to be sufficiently long, so that the two (for $S_{11}(t)$) or three (for $S_{121}(t)$) particle positions at the time of the gradient pulse applications are uncorrelated, which leads, however, to a faster convergence to the long-time limit than for method 3. When considerably reducing the diffusion time [Fig. 4(d)], the typical distance between these two or three positions of the particles gets smaller due to the occurring correlation, which results in a smaller reconstructed pore size.

The main limiting factor for the experiments conducted here was the accessible gradient amplitude of the MR scanner for the pores of small size, such as the triangles with $L = 3400$ μm. For the larger triangular pores ($L = 5750$ μm), reaching the long-time limit became challenging due to T2-relaxation.

Simple geometries such as the triangle, for which Laplacian eigenfunctions are known explicitly allowing fast and accurate simulations of the diffusion encoded signal, are a useful tool to study the influence of a varying gradient scheme on the final pore image, e.g. if no optimal settings are achievable due to hardware limitations. The demonstration experiments using other pore shapes show that pore imaging is not restricted to specific domains and that the artifacts observed for triangular domains such as edge enhancement occur in a similar manner.

Regarding applications to biological samples or porous media, besides the requirements on the gradient timing and the high required gradient amplitude, the presence of open connected geometries [36], extracellular signal components, cell membrane permeability and physiological noise result in additional challenges, which have been discussed previously in detail [16, 24]. Additionally, the short-gradient approaches (methods 1 and 2) perform worse than the long-narrow approach in the presence of broad size distributions [17].

On the other hand, the sequence design for the original long-narrow approach is very limited, because it requires at least one long gradient. This problem is non-existent for the short-gradient technique: This new flexibility is a key advantage of the short-gradient technique and may be used to employ more advanced MR sequence techniques, such as potentially stimulated echoes, to reach the long-time limit exploiting the longer $T_1$-relaxation time compared to $T_2$, which may enable imaging of larger pore sizes. While it has been experimentally demonstrated that q-space and DDE measurements yield the expected attenuation and diffraction patterns using stimulated echoes [37-39], the possibility to preserve the required phase information in DDE experiments for stimulated echoes and the corresponding appropriate choice of the RF pulses has not been investigated in detail so far to the authors' knowledge. This will have to be investigated in a future work. Additionally, short gradients may also be employed to distinguish between cases of pore size or shape distributions and cases of identical pores with inhomogeneous magnetization density inside. Unlike for the long-narrow approach, these two cases are distinguishable for the DDE approach [17]. Even if not all requirements for directly obtaining pore shapes may be sufficiently met, the additional information contained in the DDE measurements may nonetheless be valuable for a more reliable estimation of parameters of size distributions such as mean value and width.

## Acknowledgements

Financial support by the DFG (grant no. KU 3362/1-1 and LA 2804/2-1) is gratefully acknowledged.

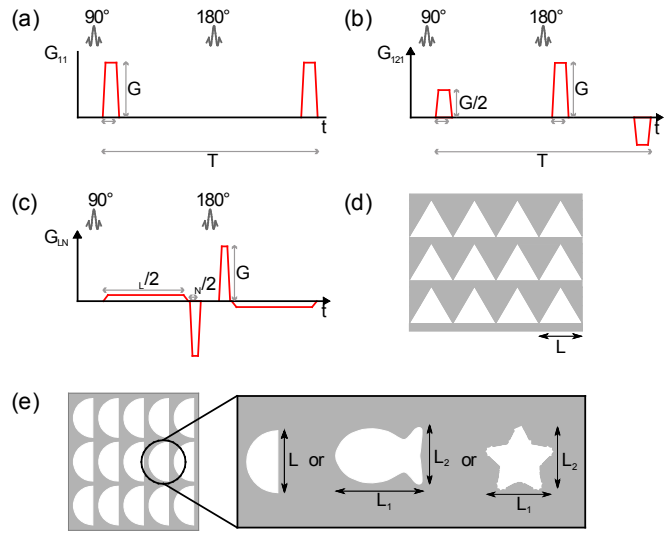

**FIG. 1.** (Color online) Gradient schemes and pore imaging phantoms used for this work. 180° RF pulses were inserted into the gradient profiles to realize spin echo sequences. (a) q-space profile $G_{11}(t)$. (b) Double diffusion encoding profile $G_{121}(t)$. (c) Long-narrow profile $G_{\text{LN}}(t)$. The gradient durations $\delta$, $\delta_{\text{N}}$ and $\delta_{\text{L}}$ are composed of the gradient flat top time plus ramp up time. In the phantoms, closed pores are aligned as an array. (d) Phantom type I: Pores of equilateral triangular shape with edge length $L$ = 3400 µm or 5750 µm. (e) Phantom type II: Pores shaped as half-moons ($L$ = 3450 µm), fish ($L_1$ = 3450 µm, $L_2$ = 2350 µm) and stars ($L_1$ = 3100 µm, $L_2$ = 3000 µm).

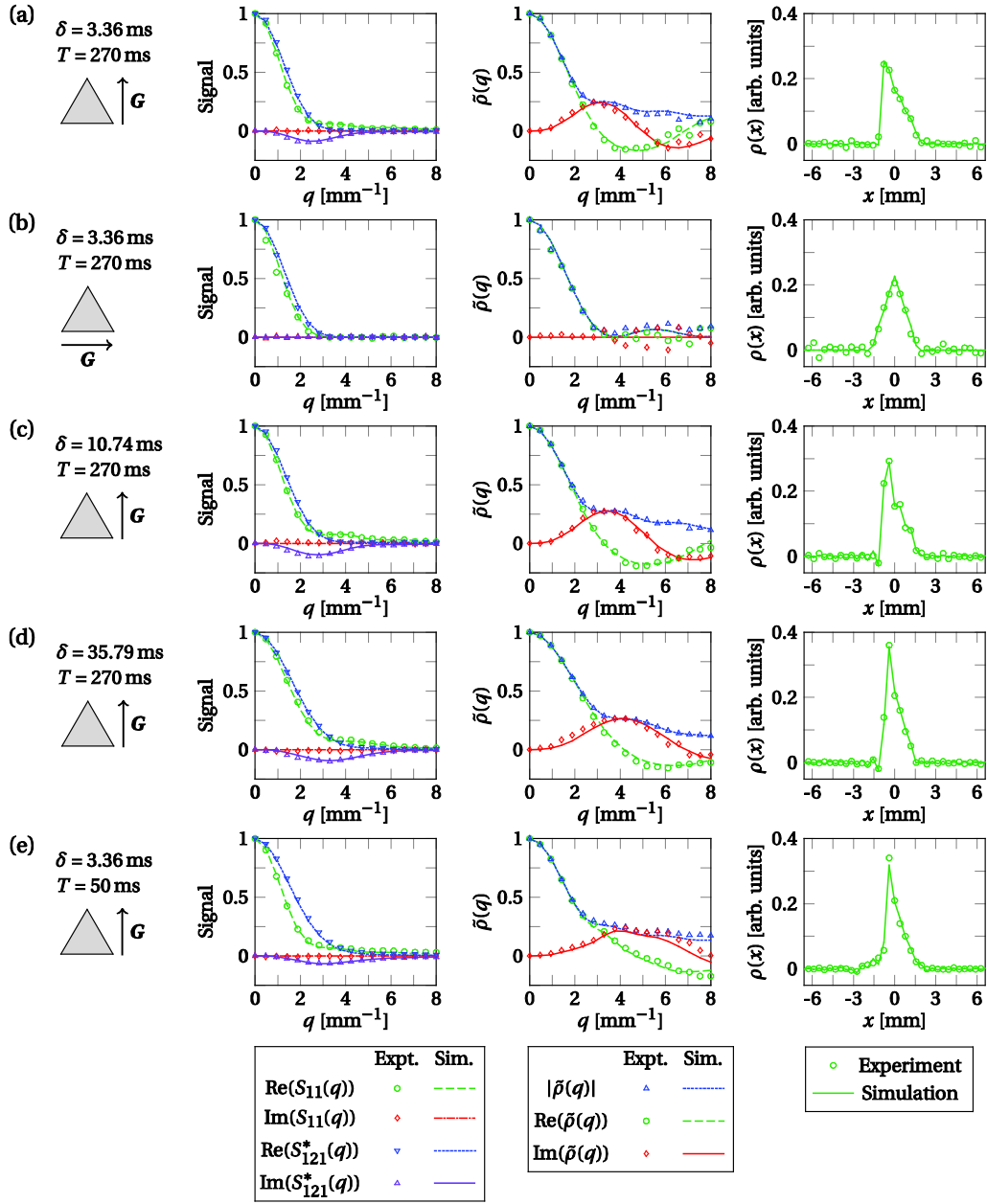

**FIG. 2.** (Color online) One-dimensional pore imaging experiments for the equilateral triangle obtained by method 1. q-Space and DDE signals were acquired radially in q-space and were used to calculate the form factor $\tilde{\rho}(q)$ followed by an inverse Fourier transform to obtain the pore space function $\rho(x)$ in x-space. The vector $\mathbf{G}$ indicates the gradient direction of the radial acquisitions. $\delta$ is the gradient pulse duration, $T$ is the diffusion time. Experimental results are indicated by symbols and simulations by lines. (a) For the vertical gradient direction, the DDE signal $S_{121}(q)$ is complex, which allows estimating the phase of $\tilde{\rho}(q)$. The magnitude of $\tilde{\rho}(q)$ is obtained from the q-space signal $S_{11}(q)$. $\rho(x)$ corresponds to the projection of the triangle domain onto the gradient direction. (b) For the horizontal gradient direction, the projection is mirror-symmetric resulting in the real DDE signal. Longer gradient

pulse durations were used in (c, d) and a shorter diffusion time was used in (e). Both entails a worse reconstruction of the pore shape. For a better visualization of the imaginary part of $S_{121}(q)$, the complex conjugate signals were plotted. The edge length of the domain was $L$ = 3400 μm. 18 q-values were measured. The maximum q-value $q_{\max}$ = 8 mm$^{-1}$ corresponds to a nominal resolution of $\Delta x$ = 0.39 mm.

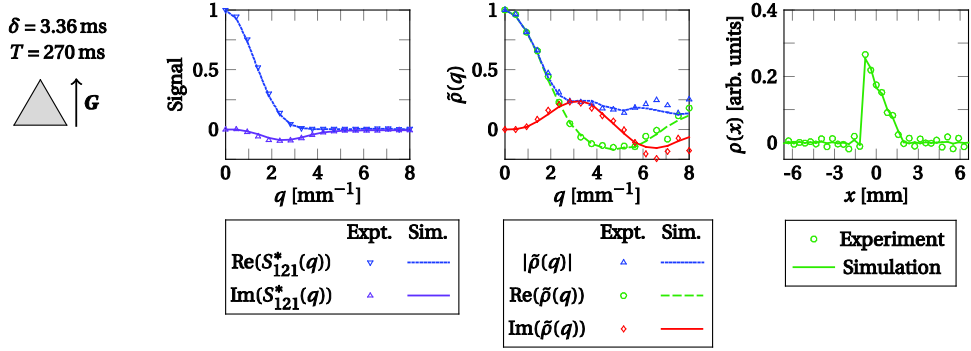

**FIG. 3.** (Color online) One-dimensional pore imaging experiment for the equilateral triangle by double diffusion encoding only (method 2) using the same phantom as in Fig. 2. The DDE signal was acquired in the vertical direction as indicated by the vector $G$ and was used to calculate both magnitude and phase of $\tilde{\rho}(q)$ with subsequent inverse Fourier transform $\rho(x)$ in x-space. Experimental results are indicated by symbols and simulations by lines. 18 q-values were measured. For better visualization of the imaginary part, the complex conjugate of $S_{121}(q)$ was plotted. Maximum q-value $q_{max}$= 8 mm$^{-1}$ corresponds to a nominal resolution of $\Delta x$ = 0.39 mm.

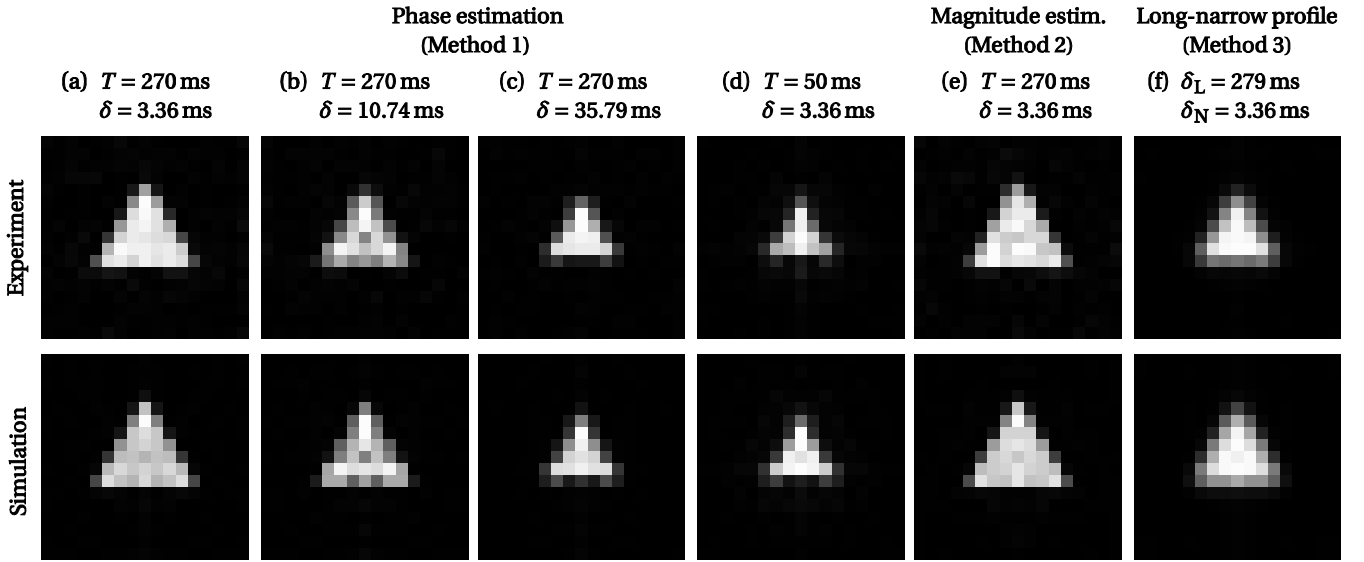

**FIG. 4.** Diffusion pore images of equilateral triangles. Experimental results in the upper row and simulations below. (a-d) Pore images obtained by method 1 with the combination of q-space and DDE-measurements and subsequent phase estimation. (e) Use of method 2: magnitude and phase estimation employing the DDE-measurement only. (f) Pore images acquired with method 3 using the long-narrow gradient profile. Additional parameters for (a) to (f): edge length $L$ = 3400 µm, 37 directions, 9 q-values per spoke, $q_{max}$ = 8 mm$^{-1}$, nominal resolution of $\Delta x$ = 0.39 mm.

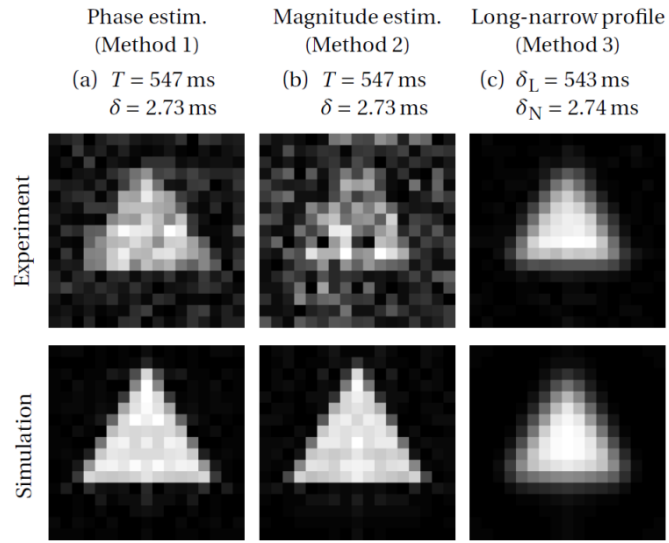

**FIG. 5.** Pore images obtained using (a) the phase estimation method (method 1), (b) the magnitude estimation method (method 2) and (c) the long-narrow approach (method 3) in a phantom with larger triangular pores (edge length $L$ = 5750 µm). Additional parameters: 37 directions, 10 q-values per spoke, $q_{max}$= 6.5 mm$^{-1}$, nominal resolution of $\Delta x$ = 0.48 mm).

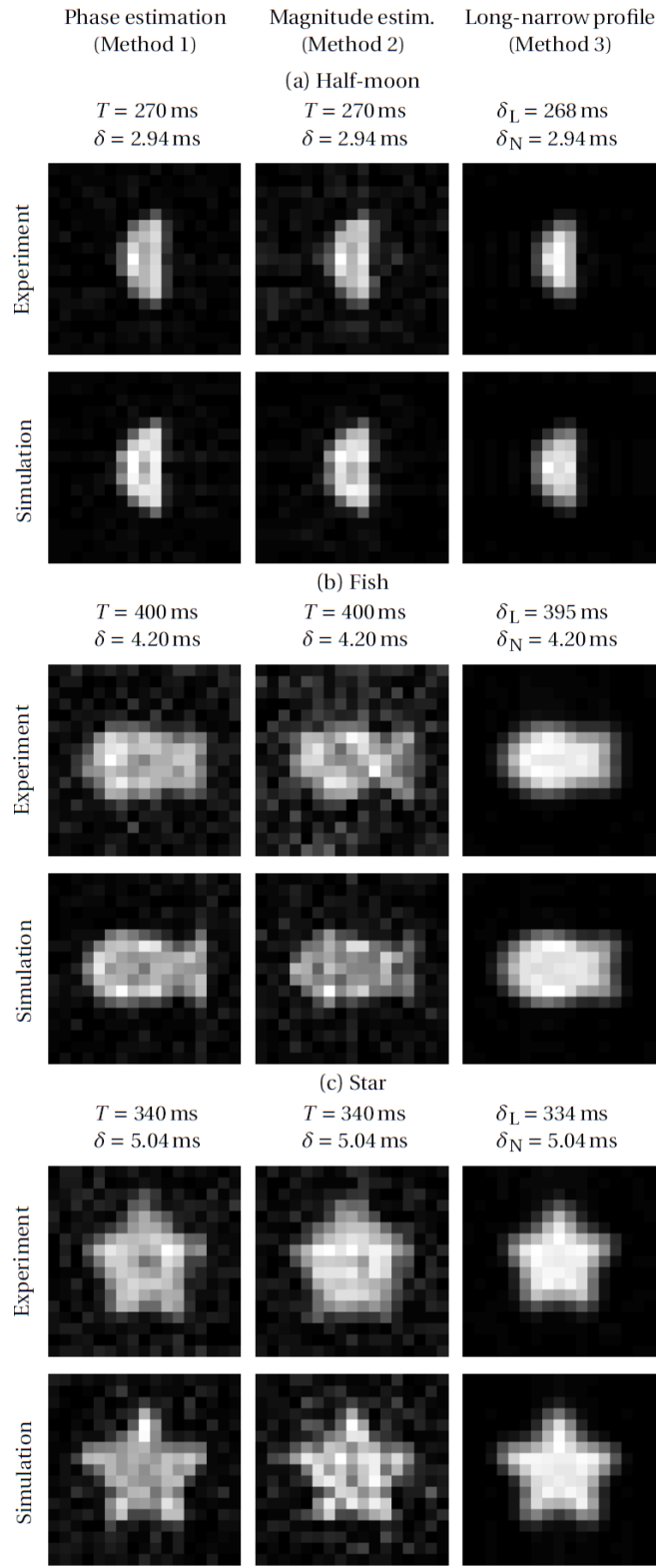

**FIG. 6.** Diffusion pore images of domains with further pore shapes. Pores were obtained using method 1 (left column), method 2 (middle column) and method 3 (right column). For each pore shape, the experimental results are shown in the upper row and simulations below. Additional Parameters: 37

directions, (a) $q_{max}$= 7 mm$^{-1}$, 8 q-values per spoke, nominal resolution of $\Delta x$ = 0.45 mm, (b) $q_{max}$= 10 mm$^{-1}$, 10 q-values, $\Delta x$ = 0.31 mm, (c) $q_{max}$= 12 mm$^{-1}$, 12 q-values, $\Delta x$ = 0.26 mm.